\documentclass{iaus}
\usepackage{graphicx}

\title[B Abundances in Diffuse Clouds]
{Boron Abundances in Diffuse Interstellar Clouds}

\author[A. M. Ritchey \etal\ ]
{A. M. Ritchey$^1$, S. R. Federman$^1$, Y. Sheffer$^1$\thanks{Present address: Department of Astronomy, University of Maryland, College Park, MD 20742, USA} \and D. L. Lambert$^2$}

\affiliation{$^1$Department of Physics and Astronomy, University of Toledo, \\
2801 West Bancroft Street, Toledo, OH 43606, USA \\
email: {\tt adam.ritchey@utoledo.edu}, {\tt steven.federman@utoledo.edu}, {\tt ysheffe@utnet.utoledo.edu} \\[\affilskip]
$^2$W.J. McDonald Observatory, University of Texas at Austin, \\
1 University Station, Austin, TX 78712, USA \\
email: {\tt dll@astro.as.utexas.edu}}

\pubyear{2010}
\volume{268}
\jname{Light Elements in the Universe}
\editors{C. Charbonnel, M. Tosi, F. Primas \& C. Chiappini, eds.}
\begin{document}

\maketitle

\begin{abstract}
We present a comprehensive survey of B abundances in diffuse interstellar clouds from \emph{HST}/STIS observations along 56 Galactic sight lines. Our sample is the result of a complete search of archival STIS data for the B~{\small II} $\lambda$1362 resonance line, with each detection confirmed by the presence of absorption from other dominant ions at the same velocity. The data probe a range of astrophysical environments including both high-density regions of massive star formation as well as low-density paths through the Galactic halo, allowing us to clearly define the trend of B depletion onto interstellar grains as a function of gas density. Many extended sight lines exhibit complex absorption profiles that trace both local gas and gas associated with either the Sagittarius-Carina or Perseus spiral arm. Our analysis indicates a higher B/O ratio in the inner Sagittarius-Carina spiral arm than in the vicinity of the Sun, which may suggest that B production in the current epoch is dominated by a secondary process. The average gas-phase B abundance in the warm diffuse ISM [log $\epsilon$(B) = $2.38\pm0.10$] is consistent with the abundances determined for a variety of Galactic disk stars, but is depleted by 60\% relative to the solar system value. Our survey also reveals sight lines with enhanced B abundances that potentially trace recent production of $^{11}$B either by cosmic-ray or neutrino-induced spallation. Such sight lines will be key to discerning the relative importance of the two production routes for $^{11}$B synthesis.
\end{abstract}

\firstsection

\section{Introduction}
Production of the two stable isotopes of boron results from spallation reactions between energetic particles and interstellar nuclei. The less abundant $^{10}$B is mainly a product of Galactic cosmic-ray (GCR) spallation (e.g., \cite[Meneguzzi, Audouze \& Reeves 1971]{mar71}; \cite[Ramaty \etal\ 1997]{ram97}), which typically involves relativistic protons and $\alpha$-particles impinging on CNO nuclei in the interstellar medium (ISM) but can also occur as accelerated CNO nuclei are spalled from ambient interstellar H and He. While a significant amount of $^{11}$B is produced through GCR spallation reactions, an additional source is required to raise the isotopic ratio $^{11}$B/$^{10}$B from the value predicted by models of cosmic-ray spallation (2.4; \cite[Meneguzzi \etal\ 1971]{mar71}) to the value measured in carbonaceous chondrites (4.0; \cite[Lodders 2003]{lod03}). The $\nu$-process, or neutrino-induced spallation in Type II supernovae, has often been invoked to account for the discrepancy in the predicted versus measured boron isotopic ratios because, while the yields for $^{11}$B are substantial, virtually no $^{10}$B is produced (\cite[Woosley \etal\ 1990]{woo90}). Without the $\nu$-process, enhanced synthesis of $^{11}$B could be attributed to an increased flux of low-energy cosmic rays, which are unobservable from the Earth due to the modulating effect of the solar wind.

All of the processes that produce boron in significant quantities occur in, or are closely associated with, the interstellar medium. Even the $^{11}$B synthesized in core collapse supernovae will quickly be injected into the surrounding interstellar gas. It therefore becomes a critical test of the theoretical ideas concerning boron nucleosynthesis to measure its interstellar abundance. The first detection of interstellar boron was reported by \cite{my80}, who used the \emph{Copernicus} satellite to measure B~{\small II} $\lambda$1362 along the line of sight to $\kappa$ Ori. They derived an interstellar abundance of log~$\epsilon$(B)~=~$2.2\pm0.2$, which was consistent with the then-current stellar value of $2.3\pm0.2$ (\cite[Boesgaard \& Heacox 1978]{bh78}), assumed to be the Galactic value. \cite{fed96a}, using GHRS on board \emph{HST}, presented the first measurement of $^{11}$B/$^{10}$B outside the solar system. Their analysis, along with later work by \cite{lam98}, showed that the solar system ratio is not anomalous but probably representative of the local Galactic neighborhood. The survey by \cite{hss00} expanded the sample of interstellar boron abundances to the extended sight lines accessible to \emph{HST}/STIS. These authors found clear evidence for boron depletion onto interstellar grains and derived a lower limit to the present-day total interstellar boron abundance of $\gtrsim$ 2.40 $\pm$ 0.13.

The discovery of newly synthesized lithium toward $o$ Per, a line of sight that passes very near to the massive star-forming region IC 348 and has a low $^7$Li/$^6$Li ratio consistent with predictions of GCR spallation (\cite[Knauth \etal\ 2000a]{kna00a}; \cite[2000b]{kna00b}), prompted us to seek $^{11}$B/$^{10}$B ratios along this and other nearby sight lines in the Per OB2 association with STIS. Ultimately, the acquired spectra toward 40~Per, $o$~Per, $\zeta$~Per, and X~Per lacked the signal-to-noise ratio required to yield meaningful results on $^{11}$B/$^{10}$B but did provide accurate B column densities. Thus, we shifted our focus to determining elemental boron abundances for a much larger Galactic sample.

\section{STIS Archival Survey}
All archival STIS datasets employing either the E140H or E140M grating were examined in an effort to find unambiguous interstellar absorption from O~{\small I} $\lambda$1355, Cu~{\small II} $\lambda$1358, and Ga~{\small II} $\lambda$1414. Subsequent searches for absorption from B~{\small II} $\lambda$1362 at the same velocity resulted in a sample of 56 Galactic sight lines. All of the above species represent the dominant ionization stage of their particular element in neutral diffuse clouds and are thus expected to coexist. As in our previous work (\cite[Federman \etal\ 1996a]{fed96a}; \cite[Lambert \etal\ 1998]{lam98}), the stronger O~{\small I}, Cu~{\small II}, and Ga~{\small II} lines serve as templates of interstellar component structure when determining B~{\small II} column densities. The STIS spectra acquired with the E140H grating are characterized by velocity resolutions in the range $\Delta v$~=~2.1$-$3.6~km~s$^{-1}$, while E140M spectra have resolutions of 6.5$-$7.9~km~s$^{-1}$. Thus, to help constrain the velocity structure along the lines of sight in our sample, we obtained high-resolution ($\Delta v$~=~1.6$-$1.8~km~s$^{-1}$) ground-based data on Ca~{\small II} $\lambda$3933 and K~{\small I} $\lambda$7698 for many directions either at McDonald Observatory or from the literature (e.g., \cite[Welty, Morton \& Hobbs 1996]{wmh96}; \cite[Welty \& Hobbs 2001]{wh01}; \cite[Pan \etal\ 2004]{pan04}). We also incorporated into our analysis five sight lines with previous measurements of interstellar B from GHRS (\cite[Jura \etal\ 1996]{jur96}; \cite[Lambert \etal\ 1998]{lam98}; \cite[Howk \etal\ 2000]{hss00}) but did not rederive abundances in these directions.

\begin{figure}[ht]
\begin{center}
\includegraphics[width=0.9\textwidth]{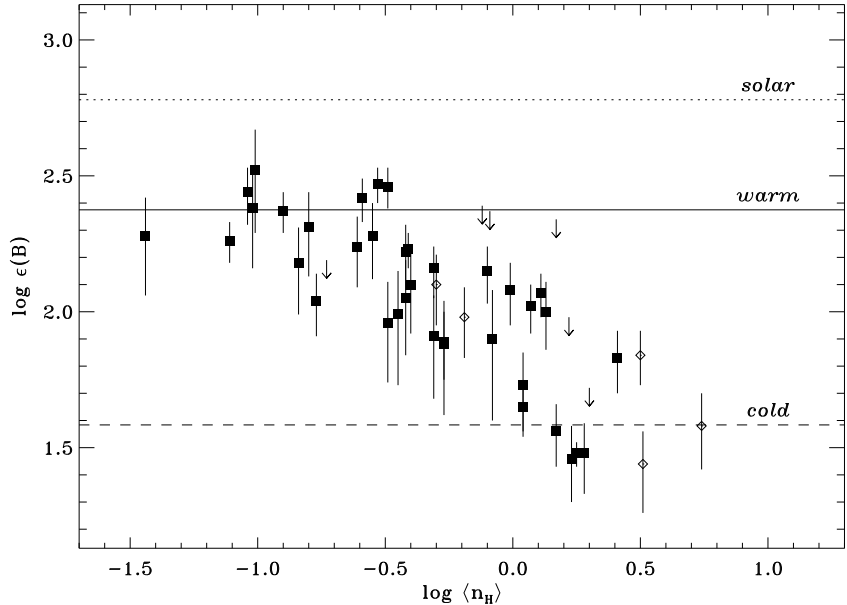} 
\caption{Gas-phase interstellar B abundances versus average line-of-sight hydrogen densities. Solid symbols (and upper limits) denote the STIS boron sample (this work). Open symbols represent GHRS measurements from the literature (see text). The dotted line indicates the solar system abundance (2.78$\pm$0.04; \cite[Lodders 2003]{lod03}), while the solid and dashed lines correspond to the mean abundances in warm gas (2.38$\pm$0.10) and cold clouds (1.58$\pm$0.16), respectively.}
\label{fig1}
\end{center}
\end{figure}

The full boron sample probes a diverse array of astrophysical environments, including both high-density regions of massive star formation (e.g., Per OB2 and Cep OB2) as well as low-density paths through the Galactic halo (e.g., the sight lines to HD~156110 and HD~187311). Some directions sample quite local gas (e.g., in Sco OB2 at $d$ = 160 pc), while others trace distant spiral arms (e.g., HD 104705, located beyond the Sagittarius-Carina arm at $d$ = 3.9 kpc). These characteristics enable a detailed investigation of the effect of physical environment on the observed abundances of boron in diffuse clouds. More information on the boron sample can be found in \cite{rit10}.

\section{Abundance Analysis}
As a first step in the analysis, the absorption profiles of O~{\small I}, Cu~{\small II}, and Ga~{\small II} were synthesized with the rms-minimizing code ISMOD (Y. Sheffer, unpublished). These fits yielded the column density, velocity, and $b$-value of each absorption component along the line of sight. The B~{\small II} line was then fitted by holding the $b$-values, relative velocities, and relative component strengths constant and varying only the absolute velocity of the profile and the total column density. The results of the O~{\small I}, Cu~{\small II}, and Ga~{\small II} fits were each applied separately to the B~{\small II} line. Additionally, if Ca~{\small II} and K~{\small I} data were available, these lines were synthesized and the results were used as a high-resolution template for synthesizing the B~{\small II} profile. Final B~{\small II} column densities were determined by taking the mean of these fits, which in all cases were mutually self-consistent.

\begin{table}[ht]
\begin{center}
\caption[Stellar Disk and Interstellar Boron Abundances]{Stellar Disk and Interstellar Boron Abundances}
\label{tab1}
\begin{tabular}{lcc}\hline
 & log $\epsilon$(B) & Reference \\
\hline
ISM (warm) & 2.38 $\pm$ 0.10 & This Work \\
G stars (Orion) & 2.35 $\pm$ 0.30 & \cite{cun00b} \\
FG stars & 2.51 $\pm$ 0.20 & \cite{cun00a} \\
B stars & 2.54 $\pm$ 0.17 & \cite{ven02} \\
Solar (photospheric) & 2.70 $\pm$ 0.17 & \cite{cs99} \\
Solar (meteoritic) & 2.78 $\pm$ 0.04 & \cite{lod03} \\
\hline
\end{tabular}
\end{center}
\end{table}

When a sight line exhibited multiple complexes of absorption components well separated in velocity, the various templates were constructed for each complex and fitted to that portion of the B~{\small II} profile independently. This technique allows for the possibility that elemental abundance ratios may vary from complex to complex along the line of sight. Indeed, we find suggestive evidence for a higher B/O ratio in components tracing the inner Sagittarius-Carina spiral arm than in those sampling local gas in the same direction. Abundances of secondary elements increase relative to those of primary ones toward the Galactic center due to enhanced rates of star formation and stellar nucleosynthesis. If confirmed, an elevated B/O ratio toward the inner Galaxy would indicate the secondary nature of boron, which in turn would cast doubt on the efficiency of the $\nu$-process, a primary production mechanism.

Under the assumption that all boron in diffuse clouds is singly ionized, the B~{\small II} column densities derived above can be considered the total boron column densities in the gas phase. Elemental boron abundances can then be determined from knowledge of the total column densities of atomic and molecular hydrogen along the line of sight. The atomic hydrogen data come mainly from the archival study of Lyman-$\alpha$ absorption by \cite{ds94}, while the majority of H$_2$ column densities were provided by \cite{she08}. In Figure~\ref{fig1}, we plot gas-phase B abundances as a function of the average line-of-sight hydrogen density, defined as $\langle n_{\mathrm{H}} \rangle$ = [$N$(H~{\small I})~+~2$N$(H$_2$)]/$d$, where $d$ is the distance to the background star. Immediately apparent is the trend of decreasing gas-phase abundance with increasing line-of-sight density, a clear signature of the depletion of boron onto interstellar dust grains. Following \cite{jss86}, we calculated mean B abundances in the warm, low-density gas and in the cold, higher-density clouds based on an idealized model of the neutral interstellar medium (\cite[Spitzer 1985]{spi85}). The analysis shows that the depletion (relative to solar) increases from $-0.40$ dex in lines of sight with the lowest density to $-1.20$ dex for the directions sampling higher concentrations of cold clouds.

\section{Discussion and Conclusions}
Our result for the mean B abundance in warm diffuse gas, log~$\epsilon$(B)~=~$2.38\pm0.10$, represents a lower limit to the total interstellar B abundance, since some depletion is expected even in the lowest-density phase of the diffuse ISM. This value agrees remarkably well with the lower limit derived by \cite[Howk \etal\ (2000)]{hss00} but is depleted by 60\% relative to the solar system (meteoritic) value of $2.78\pm0.04$ (\cite[Lodders 2003]{lod03}). The solar abundance is traditionally used as a cosmic standard against which to measure interstellar depletion, although the abundances in F and G field dwarfs of solar metallicity or hot O and B stars are also appropriate references (e.g., \cite[Snow \& Witt 1996]{sw96}; \cite[Sofia \& Meyer 2001]{sm01}). Interestingly, our interstellar B abundance for warm gas is fairly consistent with the abundances in a variety of Galactic disk stars (see Table~\ref{tab1}). At the very least, if one were to take the F and G dwarf sample of \cite{cun00a} and the sample of B-type stars in \cite{ven02} to derive a cosmic B abundance of 2.5, then B would seem to be only lightly depleted along the lowest-density interstellar sight lines. Unless the various solar and stellar abundances can be reconciled, establishing the total interstellar abundance of boron will have to await deeper insight into the gas-grain interactions responsible for removing atoms in diffuse clouds from the gas phase.

While the absolute level of boron depletion will remain controversial without a definitive cosmic standard, the variation in relative depletion has clearly been demonstrated here (Figure~\ref{fig1}). It then becomes possible to identify intrinsic variations in abundance superimposed on the general trend due to depletion. Enhanced boron abundances are expected, for example, in regions shaped by Type II supernovae, since these are the sources most likely responsible for the acceleration of Galactic cosmic rays and are also sites of the $\nu$-process. Any recent production of $^{11}$B in such a region, due either to cosmic-ray or neutrino-induced spallation, should manifest itself as a local enhancement in the total boron abundance. Depending on the degree of localization, however, the observational signature may be difficult to discern.

Nevertheless, we did find evidence for enhanced boron abundances in a few directions from our large sample of interstellar sight lines. The abundance we derive of log~$\epsilon$(B)~=~$2.47\pm0.06$ for the line of sight to HD 93222, a member of Collinder 228 in the Carina Nebula, is enhanced by 0.27 dex relative to sight lines with similar average densities and is significantly elevated compared to the values for three other sight lines in the Carina Nebula. The stars HD 93205, CPD$-$59 2603, and HDE 303308, all members of Trumpler 16, have interstellar B abundances of $2.28\pm0.12$, $2.23\pm0.06$, and $2.05\pm0.14$, respectively, and lie just 23$^{\prime}$ to the north of HD 93222. \cite{wal07} discuss high-velocity expanding structures seen in interstellar absorption lines toward many of these cluster members in the context of a supernova remnant (SNR) in this direction. They note that the highest known interstellar velocities in the nebula occur in the spectrum of HD 93222.

For HD 43818, a member of the Gem OB1 association, we derive a B abundance of $2.07\pm0.07$. The line of sight to this star is characterized by a factor of 4 higher average density than those in Carina. Since higher depletion is therefore expected, the abundance in this direction represents an enhancement, which is found to be 0.26 dex over similarly dense sight lines. The proximity of this line of sight to IC 443, a young SNR known to be interacting with nearby molecular gas, suggests a possible nucleosynthetic origin for the enhancement. The presence of hadronic cosmic rays accelerated by the SNR and their interaction with ambient molecular material was revealed by VERITAS observations of very-high energy $\gamma$-ray emission (\cite[Acciari \etal\ 2009]{acc09}). HD 43818, however, lies considerably to the north of the $\gamma$-ray source and so may not be related. We are currently pursuing $^7$Li/$^6$Li ratios and Li and Rb abundances toward stars closer to IC~443 with the Hobby-Eberly Telescope at McDonald Observatory in an effort to constrain the contribution from massive stars to the synthesis of these elements.

Finally, the line of sight to $o$ Per, which is located just 8$^{\prime}$ to the north of the star-forming region IC 348, has a B abundance enhanced by 0.18 dex relative to the other three sight lines in Per OB2. While the enhancement is only modest (50\%), it becomes significant in light of the fact that the other sight lines show very little scatter in log~$\epsilon$(B). For 40~Per, $\zeta$~Per, and X~Per, we derive abundances of $1.48\pm0.11$, $1.46\pm0.12$, and $1.48\pm0.04$, while for $o$ Per we find an abundance of $1.65\pm0.09$. Considering the low $^7$Li/$^6$Li ratio in this direction (\cite[Knauth \etal\ 2000a]{kna00a}; \cite[2000b]{kna00b}; \cite[2003]{kfl03}) and an enhanced cosmic-ray flux, which was inferred from measurements of interstellar OH (\cite[Federman, Weber \& Lambert 1996b]{fwl96b}) and is consistent with an upper limit derived from observations of H$_3^+$ (\cite [Indriolo \etal\ 2007]{ind07}), evidence seems to be mounting of the effect of cosmic-ray spallation reactions on the interstellar abundances of Li and B near IC 348. Recently, Li data were obtained toward the fainter stars of IC 348, itself. Complementary data on interstellar B should now be acquired for these stars, perhaps with the Cosmic Origins Spectrograph, so that the abundance enhancements resulting from cosmic-ray interactions with interstellar clouds can be traced in more detail. In this way, a clearer picture of light element nucleosynthesis will emerge.

\acknowledgments
This research was funded by the Space Telescope Science Institute (STScI) through grant HST-AR-11247.01-A. The data were obtained from the Multimission Archive at STScI, operated by the Association of Universities for Research in Astronomy, Inc. under NASA contract NAS5-26555. A. M. R. would like to thank the Swiss National Science Foundation for a grant provided to cover attendance at the IAU Symposium 268.





\begin{thebibliography}{}
\bibitem[Acciari \etal\ (2009)]{acc09}
{Acciari, V. A., Aliu, E., Arlen, T., \& VERITAS Collaboration.} 2009,
\textit{ApJ}, 698, L133

\bibitem[Boesgaard \& Heacox (1978)]{bh78}
{Boesgaard, A. M., \& Heacox, W. D.} 1978,
\textit{ApJ}, 226, 888

\bibitem[Cunha \& Smith (1999)]{cs99}
{Cunha, K., \& Smith, V. V.} 1999,
\textit{ApJ}, 512, 1006

\bibitem[Cunha \etal\ (2000a)]{cun00a}
{Cunha, K., Smith, V. V., Boesgaard, A. M., \& Lambert, D. L.} 2000a,
\textit{ApJ}, 530, 939

\bibitem[Cunha \etal\ (2000b)]{cun00b}
{Cunha, K., Smith, V. V., Parizot, E., \& Lambert, D. L.} 2000b,
\textit{ApJ}, 543, 850

\bibitem[Diplas \& Savage (1994)]{ds94}
{Diplas, A., \& Savage, B. D.} 1994,
\textit{ApJS}, 93, 211

\bibitem[Federman \etal\ (1996a)]{fed96a}
{Federman, S. R., Lambert, D. L., Cardelli, J. A., \& Sheffer, Y.} 1996a,
\textit{Nature}, 381, 764

\bibitem[Federman, Weber \& Lambert (1996b)]{fwl96b}
{Federman, S. R., Weber, J., \& Lambert, D. L.} 1996b,
\textit{ApJ}, 463, 181

\bibitem[Howk, Sembach \& Savage (2000)]{hss00}
{Howk, J. C., Sembach, K. R., \& Savage, B. D.} 2000,
\textit{ApJ}, 543, 278

\bibitem[Indriolo \etal\ (2007)]{ind07}
{Indriolo, N., Geballe, T. R., Oka, T., \& McCall, B. J.} 2007,
\textit{ApJ}, 671, 1736

\bibitem[Jenkins, Savage \& Spitzer (1986)]{jss86}
{Jenkins, E. B., Savage, B. D., \& Spitzer, L.} 1986,
\textit{ApJ}, 301, 355

\bibitem[Jura \etal\ (1996)]{jur96}
{Jura, M., Meyer, D. M., Hawkins, I., \& Cardelli, J. A.} 1996,
\textit{ApJ}, 456, 598

\bibitem[Knauth \etal\ (2000a)]{kna00a}
{Knauth, D. C., Federman, S. R., Lambert, D. L., \& Crane, P.} 2000a,
\textit{Nature}, 405, 656

\bibitem[Knauth \etal\ (2000b)]{kna00b}
{Knauth, D. C., Federman, S. R., Lambert, D. L., \& Crane, P.} 2000b,
in: L. da Silva, M. Spite \& J. R. de Medeiros (eds.),
\textit{The Light Elements and Their Evolution},
Proc. IAU Symposium No. 198 (San Francisco: ASP), p.\ 338

\bibitem[Knauth \etal\ (2003)]{kfl03}
{Knauth, D. C., Federman, S. R., \& Lambert, D. L.} 2003,
\textit{ApJ}, 586, 268

\bibitem[Lambert \etal\ (1998)]{lam98}
{Lambert, D. L., Sheffer, Y., Federman, S. R., Cardelli, J. A., Sofia, U. J., \& Knauth, D. C.} 1998,
\textit{ApJ}, 494, 614

\bibitem[Lodders (2003)]{lod03}
{Lodders, K.} 2003,
\textit{ApJ}, 591, 1220

\bibitem[Meneguzzi, Audouze \& Reeves (1971)]{mar71}
{Meneguzzi, M., Audouze, J., \& Reeves, H.} 1971,
\textit{A\&A}, 15, 337

\bibitem[Meneguzzi \& York (1980)]{my80}
{Meneguzzi, M., \& York, D. G.} 1980,
\textit{ApJ}, 235, L111

\bibitem[Pan \etal\ (2004)]{pan04}
{Pan, K., Federman, S. R., Cunha, K., Smith, V. V., \& Welty, D. E.} 2004,
\textit{ApJS}, 151, 313

\bibitem[Ramaty \etal\ (1997)]{ram97}
{Ramaty, R., Kozlovsky, B., Lingenfelter, R. E., \& Reeves, H.} 1997,
\textit{ApJ}, 488, 730

\bibitem[Ritchey \etal\ (2010)]{rit10}
{Ritchey, A. M., Federman, S. R., Sheffer, Y., \& Lambert, D. L.} 2010,
\textit{in preparation}

\bibitem[Sheffer \etal\ (2008)]{she08}
{Sheffer, Y., Rogers, M., Federman, S. R., Abel, N. P., Gredel, R., Lambert, D. L., \& Shaw, G.} 2008,
\textit{ApJ}, 687, 1075

\bibitem[Snow \& Witt (1996)]{sw96}
{Snow, T. P., \& Witt, A. N.} 1996,
\textit{ApJ}, 468, L65

\bibitem[Sofia \& Meyer (2001)]{sm01}
{Sofia, U. J., \& Meyer, D. M.} 2001,
\textit{ApJ}, 554, L221

\bibitem[Spitzer (1985)]{spi85}
{Spitzer, L.} 1985,
\textit{ApJ}, 290, L21

\bibitem[Venn \etal\ (2002)]{ven02}
{Venn, K. A., Brooks, A. M., Lambert, D. L., Lemke, M., Langer, N., Lennon, D. J., \& Keenan, F. P.} 2002,
\textit{ApJ}, 565, 571

\bibitem[Walborn \etal\ (2007)]{wal07}
{Walborn, N. R., Smith, N., Howarth, I. D., Kober, G. V., Gull, T. R., \& Morse, J. A.} 2007,
\textit{PASP}, 119, 156

\bibitem[Welty \& Hobbs (2001)]{wh01}
{Welty, D. E., \& Hobbs, L. M.} 2001,
\textit{ApJS}, 133, 345

\bibitem[Welty, Morton \& Hobbs (1996)]{wmh96}
{Welty, D. E., Morton, D. C., \& Hobbs, L. M.} 1996,
\textit{ApJS}, 106, 533

\bibitem[Woosley \etal\ (1990)]{woo90}
{Woosley, S. E., Hartmann, D. H., Hoffman, R. D., \& Haxton, W. C.} 1990,
\textit{ApJ}, 356, 272

\end{thebibliography}
\end{document}